\documentclass[floatfix,twocolumn,showpacs,preprintnumbers,amsmath,amssymb,pra,superscriptaddress,longbibliography]{revtex4-1}
\usepackage{color}
\usepackage[usenames,dvipsnames,svgnames,table]{xcolor}
\usepackage[colorlinks=true,linkcolor=blue,urlcolor=blue,citecolor=blue]{hyperref}
\usepackage{mathtools}
\usepackage{graphicx}
\usepackage{dcolumn}
\usepackage{array}
\usepackage{makecell}
\usepackage{lipsum}
\usepackage{bm}
\usepackage{subfigure}
\usepackage{fontenc}
\usepackage{amssymb}
\usepackage{multirow}
\usepackage{tabularx}
\usepackage{amsmath}
\usepackage{braket}
\graphicspath{{plots/}}
 \usepackage{lipsum}
\usepackage{mathrsfs}

\newcommand{\beq}{\begin{equation}}
\newcommand{\eeq}{\end{equation}}
\newcommand{\bea}{\begin{eqnarray}}
\newcommand{\eea}{\end{eqnarray}}

\begin{document}
\title{
Prediction of a roton-type feature in 
warm dense hydrogen
}

\author{Paul Hamann}
\author{Linda Kordts}
\author{Alexey Filinov}
\author{Michael Bonitz}

\affiliation{Institut f\"ur Theoretische Physik und Astrophysik, Christian-Albrechts-Universit\"at zu Kiel, D-24098 Kiel, Germany}

\author{Tobias Dornheim}

\affiliation{Center for Advanced Systems Understanding (CASUS), D-02826 G\"orlitz, Germany}
\affiliation{Helmholtz-Zentrum Dresden-Rossendorf (HZDR), D-01328 Dresden, Germany}



\author{Jan Vorberger}\email{j.vorberger@hzdr.de}
\affiliation{Helmholtz-Zentrum Dresden-Rossendorf (HZDR), D-01328 Dresden, Germany}


\begin{abstract}
In a recent Letter [T. Dornheim \textit{et al.}, Phys. Rev. Lett. \textbf{121}, 255001 (2018)], it was predicted on the basis of \textit{ab initio} quantum Monte Carlo simulations that, in a uniform electron gas, the peak $\omega_0$ of the dynamic structure factor $S(q,\omega)$ exhibits an unusual non-monotonic wave number dependence, where $d\omega_0/dq < 0$, at intermediate $q$, under strong coupling conditions. This effect was subsequently explained by the pair alignment of electrons 
[T. Dornheim \textit{et al.}, Comm. Phys. \textbf{5}, 304 (2022)].
Here we predict that this non-monotonic dispersion resembling the roton-type behavior known from superfluids should be observable in a dense, partially ionized hydrogen plasma. Based on a combination of path integral Monte Carlo simulations and linear response results for the density response function, we present the approximate range of densities,  temperatures and wave numbers and make predictions for possible experimental observations. 

\end{abstract}

\maketitle

\section{Introduction}\label{s:intro}

Over the last decades, there has been a surge of interest in the properties of matter at extreme temperatures ($T\sim10^3-10^7\,$K) and pressures ($P\sim1-10^4\,$Mbar). Such conditions play an important role in astrophysics~\cite{drake2018high,fortov_review} and occur naturally, for example in giant planet interiors~\cite{Benuzzi_Mounaix_2014,militzer1}, brown dwarfs~\cite{saumon1,becker}, and the outer layer of neutron stars~\cite{Chamel2008}. In addition, such extreme states are important for a number of practical applications, with inertial confinement fusion~\cite{hu_ICF,Betti2016,Zylstra2022} being a case in point. 
Other technological applications include the discovery of novel materials~\cite{Kraus2016,Lazicki2021} and hot-electron chemistry~\cite{Brongersma2015}.

A particularly important parameter regime is given by so-called \emph{warm dense matter} (WDM)~\cite{new_POP,wdm_book}, which is typically defined by two characteristic parameters that are of the order of one simultaneously~\cite{ott_epjd18}: 1) the density parameter $r_s=d/a_\textnormal{B}$ is given by the Wigner-Seitz radius in units of the Bohr radius, and 2) the degeneracy temperature $\Theta=k_\textnormal{B}T/E_\textnormal{F}$ where $E_\textnormal{F}$ is the usual Fermi energy of the electrons~\cite{quantum_theory}. The condition $r_s\sim\Theta\sim1$ implies that WDM exhibits an intriguing interplay of physical effects such as quantum degeneracy and diffraction, moderate to strong Coulomb coupling, and thermal excitations. The rigorous understanding of WDM thus poses a formidable challenge for theory and experiment alike.

In the laboratory, WDM can be realized using a plethora of different techniques; see Ref.~\cite{falk_wdm} for a review article.
At the same time, the diagnostics of experiments with WDM is notoriously difficult due to the extreme conditions. Indeed, often even basic parameters such as the temperature $T$, number density $n$, or the effective ionization degree $\alpha$ or mean charge per atom cannot be directly measured and have to be inferred indirectly from other observations. In this regard,  X-ray Thomson scattering (XRTS) represents a key method~\cite{siegfried_review,sheffield2010plasma,kraus_xrts,Dornheim_T_2022,Dornheim_T2_2022}. The measured scattering intensity is given by the convolution of the combined source and instrument function $R(\omega)$ with the dynamic structure factor (DSF) $S(\mathbf{q},\omega)$~\cite{siegfried_review,Dornheim_T2_2022},
\begin{eqnarray}\label{eq:I}
I(\mathbf{q},\omega) = S(\mathbf{q},\omega) \circledast R(\omega)\ ,
\end{eqnarray}
with $\mathbf{q}$ denoting the momentum transfer vector that is being determined by the scattering angle $\theta_s$ whereas $\omega$ is the corresponding energy loss. 
In practice, $R(\omega)$ is often accurately obtained by additional source monitoring at modern X-ray free-electron laser (XFEL) facilities such as the European XFEL~\cite{Tschentscher_2017} in Germany or the LCLS~\cite{LCLS_2016} in the USA, or from the characterization of backlighter sources~\cite{MacDonald_POP_2022} as they are employed for example at the National Ignition Facility~\cite{Moses_NIF}.

The DSF $S(\mathbf{q},\omega)$ is a key property in quantum many-body theory~\cite{Dornheim_review} and, in principle, contains the full thermodynamic information about the given system. Unfortunately, the numerical deconvolution that is required to solve Eq.~(\ref{eq:I}) for the DSF is highly unstable. To interpret an XRTS experiment one, therefore, has to construct a model $S_\textnormal{model}(\mathbf{q},\omega)$ which, after being convolved with $R(\omega)$, can be compared with the experimental observation.
On the one hand, this procedure, in principle, allows one to extract system parameters by determining the set of free parameters (e.g.~$T$, $n$, etc.) that result in the best fit to the experiment~\cite{siegfried_review,kraus_xrts}. On the other hand, the interpretation of the experiment then depends on the particular model; typical assumptions include the decomposition into \emph{bound} and \emph{free} electrons within the Chihara decomposition~\cite{Chihara_1987,Gregori_PRE_2003}
or adiabatic approximations for the exchange--correlation kernel in more sophisticated linear-response time-dependent density functional theory (DFT) calculations~\cite{Schoerner_arxiv_2023,Moldabekov_arxiv_2023,Dornheim_review}.

This unsatisfactory situation reflects the notorious difficulty to find a thorough theoretical description of real WDM systems as it has been explained above~\cite{new_POP,wdm_book}. This challenge has been met recently for the somewhat simplified case of a uniform electron gas (UEG)~\cite{review,quantum_theory}. More specifically, Dornheim \emph{et al.}~\cite{dornheim_dynamic,dynamic_folgepaper,Dornheim_PRE_2020} have presented the first highly accurate results for the DSF of the UEG based on a combination of extensive \emph{ab initio} path integral Monte Carlo (PIMC) simulations~\cite{dornheim_sign_problem} and the stochastic sampling of the dynamic local field correction. 
Interestingly, they have found a non-monotonic dependence of the position of the maximum in the DSF, $\omega_0(q)$, for intermediate wave numbers $q=|\mathbf{q}|$ -- an unusual behavior that resembles the well-known \emph{roton feature} in the dispersion of quantum liquids such as ultracold helium~\cite{griffin1996bose,Trigger,Godfrin2012,Nava_PRB_2013,Ferre_PRB_2016,Dornheim_SciRep_2022} as well as the plasmon dispersion of strongly coupled classical plasmas \cite{kalman_epl_10}.

\begin{figure}
    \centering
    \includegraphics[width=0.3\textwidth]{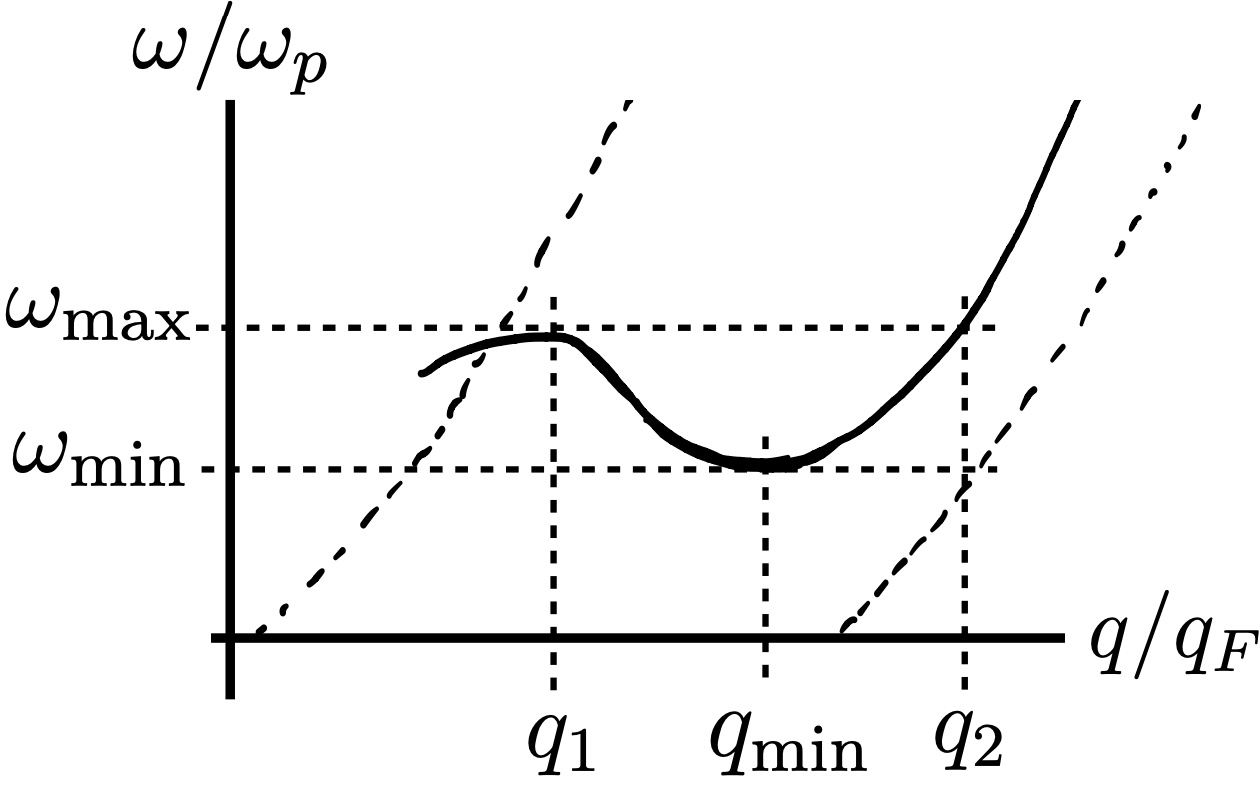}
    \caption{Sketch of the negative dispersion [peak of $S(q,\omega)$] and notation for characteristic $q$- and $\omega$-values.
At $r_s = 10$ and $\theta = 1.0$, Ref.~\cite{dornheim_prl_18} predicts $q_{\text{1}} \approx 1.1 \,  q_F $, $q_{\text{min}} \approx 1.8 \,  q_F $, $\omega_{\text{max}} \approx 1.25 \,  \omega_p $, and $\omega_{\text{min}} \approx 0.90 \,  \omega_p $, with $\omega_p= [n_e e^2/(\epsilon_0 m_e) ]^{1/2}$ denoting the electron plasma frequency. The difference $\omega_{\text{max}} - \omega_{\text{min}}$ amounts to 0.52 eV  for the uniform electron gas model.
    }
    \label{fig:dispersion_schematic}
\end{figure}

This effect is schematically illustrated in Fig.~\ref{fig:dispersion_schematic}, where we show $\omega_0(q)$ (solid black line) for $r_s=10$ at the electronic Fermi temperature, $\Theta=1$ [the relation to the physical temperature in hydrogen will be discussed below, see e.g. Fig.~\ref{fig:densShift}].
 Starting with the collective plasmon excitation around the plasma frequency $\omega_\textnormal{p}$ for $q\to0$, the frequency $\omega_0(q)$ increases with increasing $q$  and attains a local maximum at $q_1$, followed by a minimum at $q_\textnormal{min}$. The quadratic increase of $\omega_0(q)$, for $q\gg q_\textnormal{min}$ then follows from the well-known single particle dispersion $\omega\sim q^2/2$, e.g. \cite{zhandos_pop18,hamann_cpp_20}.
From a physical perspective, the non-monotonic behaviour of $\omega_0(q)$ occurs when the wavelength of the oscillation is comparable to
the mean inter-particle distance, $\lambda=2\pi/q\sim d$. Indeed, the observed reduction in the energy of a density fluctuation has been explained in Ref.~\cite{Dornheim_Nature_2022} by the alignment of pairs of electrons, leading to a decrease in the interaction energy when $\lambda\sim d$.
For completeness, we note that an alternative explanation has been given in Refs.~\cite{Takada_PRB_2016,Koskelo}, where the minimum in $\omega(q)$ has been interpreted as an excitonic mode.

\begin{figure}
\center
\includegraphics[width=0.5\textwidth]{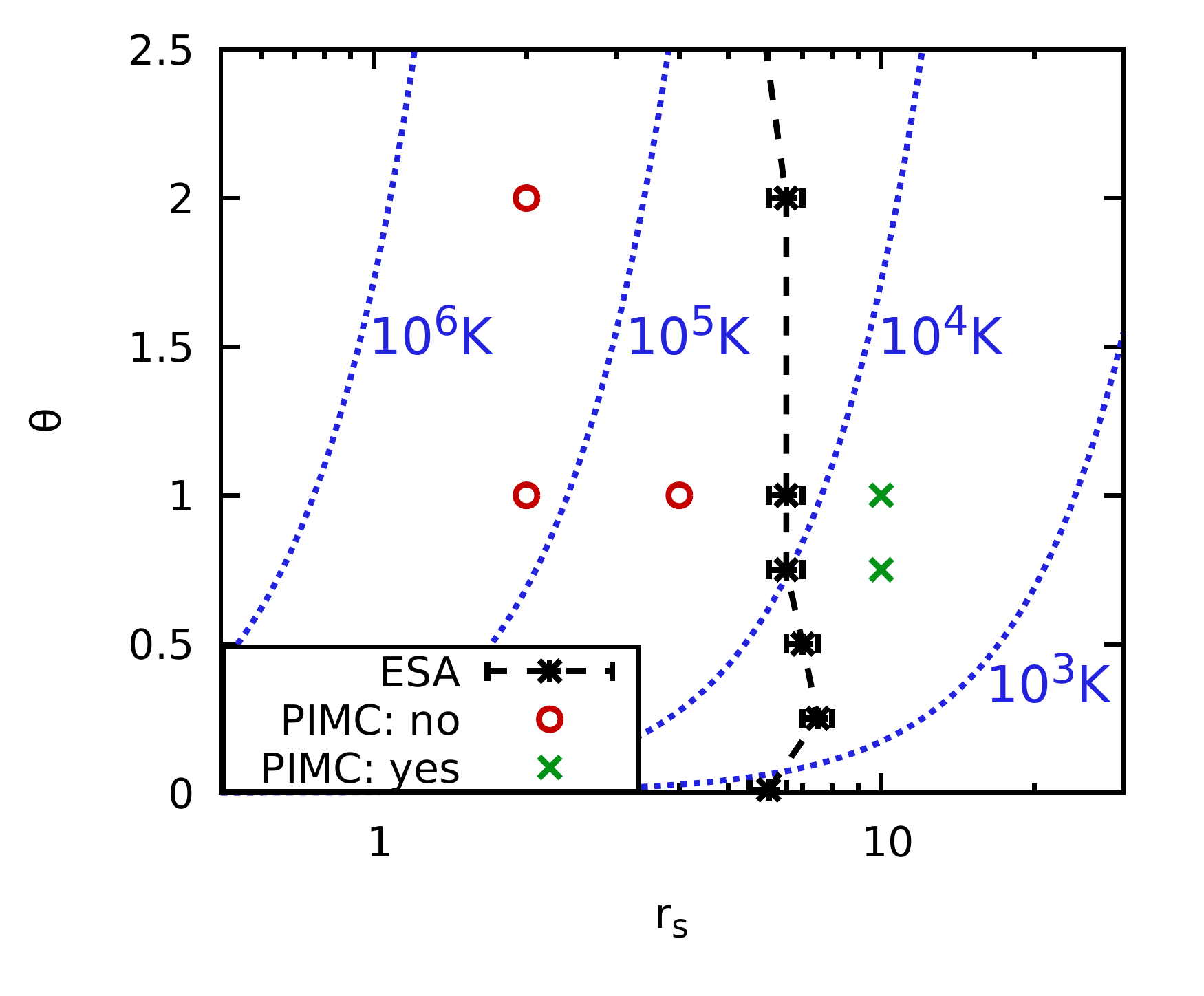}
\caption{\label{fig:phase_diagram}
Phase diagram for the predicted negative dispersion of the peak, $\omega_0(q)$, of the dynamic structure factor, $S(q,\omega)$, of the uniform electron gas in the WDM regime. Green (red) symbols: observed (not observed) negative dispersion in PIMC simulations.  ESA: predictions based on effective static approximation for the uniform electron gas of Ref.~\cite{Dornheim_PRL_2020_ESA}. The negative dispersion, $d\omega_0(q)/dq < 0$, is predicted to exist  to the right of the dashed black line, see also Fig.~\ref{fig:dispersion_schematic}.
}
\end{figure}

An additional interesting question is given by the phase diagram of the roton-type feature of the electron gas, which is shown in Fig.~\ref{fig:phase_diagram}. We observe the non-monotonic behaviour that was discussed above at sufficiently strong coupling (large $r_s$), i.e. to the right of the dashed grey line, which has been computed within the effective static approximation introduced in Refs.~\cite{Dornheim_PRL_2020_ESA,Dornheim_PRB_ESA_2021}. This prediction is consistent with the red circles and green crosses that show PIMC results based on the full dynamic local field correction. 
Being an exchange--correlation effect, the non-monotonic feature in $\omega_0(q)$ thus only occurs at sufficiently low density, i.e., strong coupling, where correlation effects are important.

While these results for the UEG are interesting, the question whether this roton-type feature will also manifest in real WDM systems 
and whether it could be detected in XRTS measurements has remained open until now.
Therefore, in this work, we extend the previous theoretical considerations to two-component systems. This requires, first, to include electron--ion collisions, which is done via the Mermin dielectric function~\cite{mermin_prb_70}. In addition, we take into account that  bound states will form at strong coupling and thus the system will only be partially ionized~\cite{schlanges-etal.95cpp,filinov_jetpl_00,QStatNonideal}. This will lead to a reduction of the number of free electrons that can participate in plasma oscillations. To investigate the existence of the roton feature in two-component plasmas, we focus on the case of hydrogen using restricted PIMC data by Militzer and Ceperley~\cite{MC01} for the degree of ionization.

Interestingly, our analysis shows that the roton feature does not only persist in hydrogen but is even
 substantially stabilized by the presence of the ions. This means, it is predicted to occur at significantly higher densities as compared to the UEG. In addition, we investigate the optimal range of temperatures and densities, and discuss the required wavenumbers and scattering angles for XRTS measurements. Our findings indicate that the effect should be resolvable in upcoming experiments with hydrogen jets~\cite{Zastrau} at modern XFEL facilities such as the European XFEL~\cite{Tschentscher_2017}.

The paper is organized as follows: In Sec.~\ref{s:ueg-data}, we introduce the required theoretical background, starting with linear-response theory and its connection to the DSF in Sec.~\ref{ss:dsf}, and a concise discussion of previous PIMC-based results for the roton-feature of the UEG in Sec.~\ref{ss:pimc}. After this, in Sec.~\ref{s:hydrogen} we analyze the plasmon dispersion for dense hydrogen and confirm the existence of a roton-type feature. There we predict the density-temperature range where the latter should be observable. We conclude with a discussion of the results in Sec.~\ref{s:discussion}.


\section{Theoretical background and previous results for the UEG}\label{s:ueg-data}
\subsection{Density response and dynamic structure factor}\label{ss:dsf}
The dynamic structure factor is related to the imaginary part of the density response function, $\chi(q,\omega)$, via the fluctuation-dissipation theorem,
    \begin{equation}
            S(q,\omega) = - \frac{\Im \,\chi(q,\omega)}{\pi n (1 - e^{-\beta\omega})}\,.
            \label{eq:dsf-fdt}
    \end{equation}
For the special case of the uniform electron gas (jellium, J), the density response function is given by~\cite{kugler1}    
    \begin{equation}
    \chi^{\rm J}(q,\omega) = \frac{\chi_0(q,\omega)}{1-v_q [1-G(q,\omega)] \chi_0(q,\omega)}\,,
    \label{eq:chi-ueg}
\end{equation}
where $v_q$ is the Fourier transform of the Coulomb potential, and $\chi_0$ is the density response of the ideal UEG. All electron-electron interaction effects beyond mean field are contained in the dynamic local field correction $G(q,\omega)$. For $G\to 0$, we recover the mean field (RPA) result. It is often sufficient to consider the static limit of the local field correction, $G(q)=G(q,0)$. This \emph{static approximation} has been shown to be highly accurate for high to moderate densities, $r_s\lesssim 4$, and has been explored in more detail in Refs.~\cite{Dornheim_PRL_2020_ESA,Dornheim_PRB_ESA_2021}.
Accurate results for $G(q,\omega)$ were recently obtained from ab initio PIMC simulations in Ref.~\cite{dornheim_prl_18}, for details see Sec.~\ref{ss:pimc}.  

Compared to the UEG model, realistic plasmas require to take into account several additional effects, in addition to purely electronic correlations. As was discussed in the introduction, the first is the scattering of electrons with ions~\cite{Bohme_PRL_2022}. This can be accounted for via the Mermin dielectric function, cf. Sec.~\ref{ss:merin}. Secondly, in two-component plasmas, electrons and ions can form bound states, thereby reducing the number of free electrons which are participating in collective plasma oscillations and eventually contribute to the roton feature. This will be analyzed in Sec.~\ref{s:hydrogen}.

\subsection{PIMC results for the UEG}\label{ss:pimc}
\begin{figure*}
    \centering
    \includegraphics[width=0.43\textwidth]{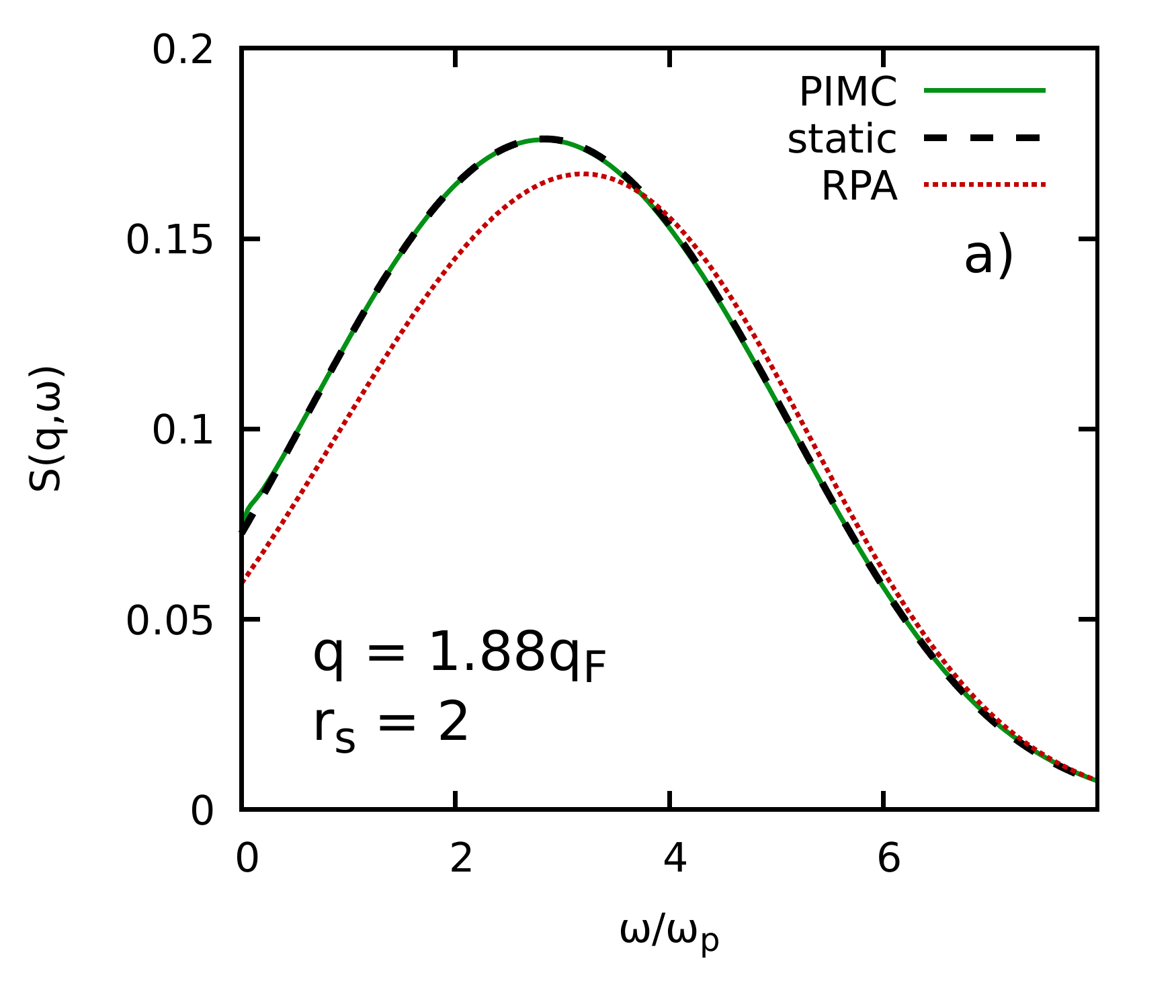}\includegraphics[width=0.43\textwidth]{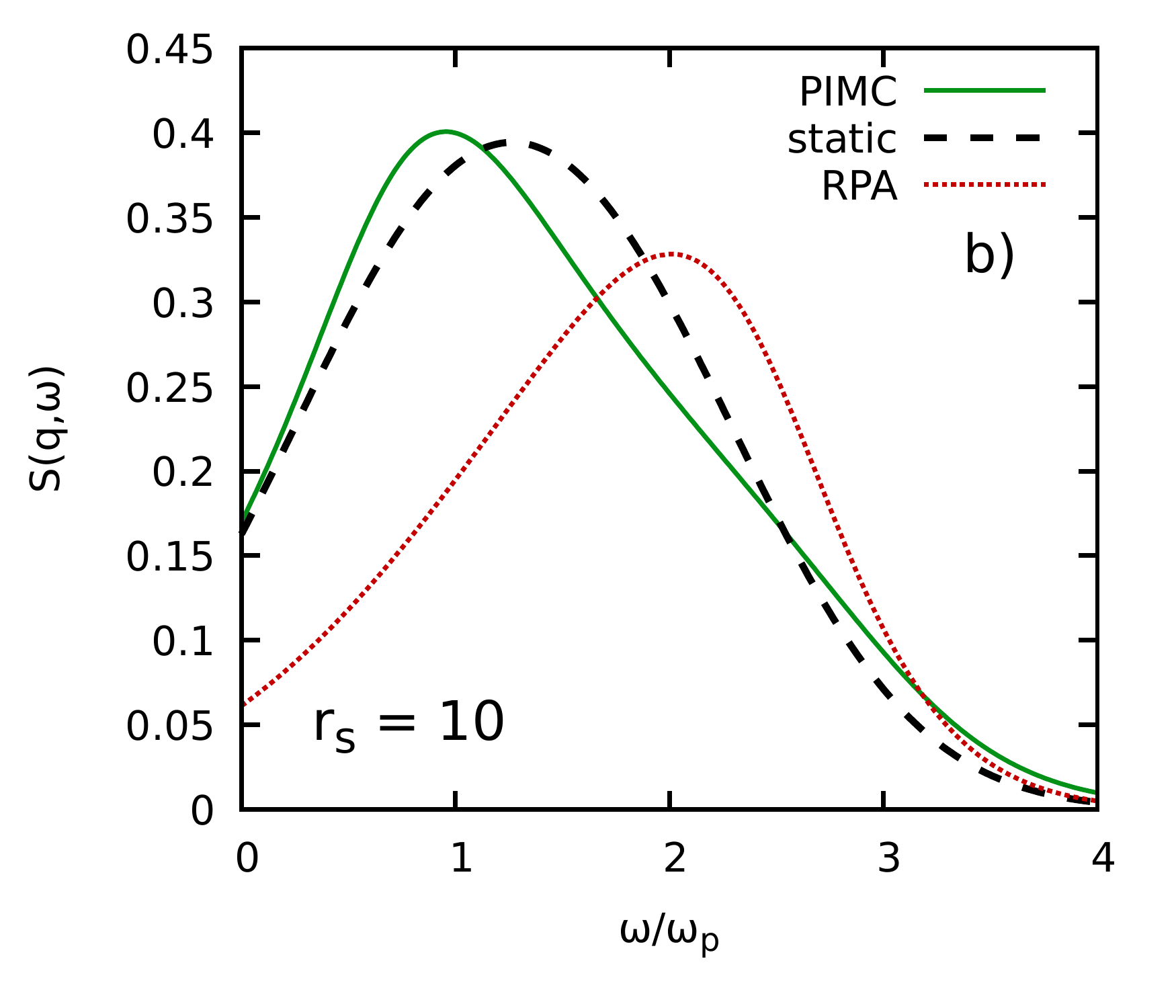}\\\includegraphics[width=0.43\textwidth]{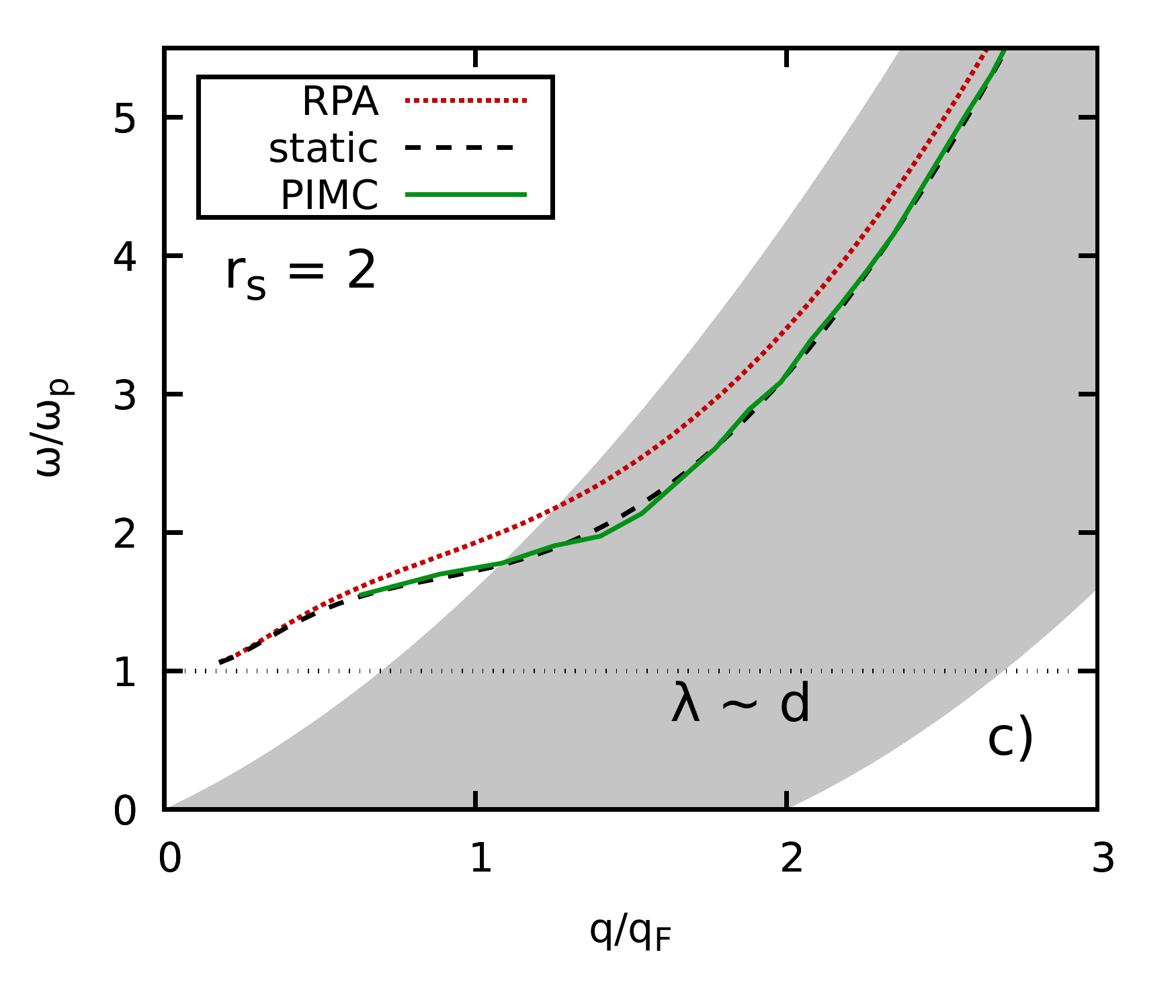}\includegraphics[width=0.43\textwidth]{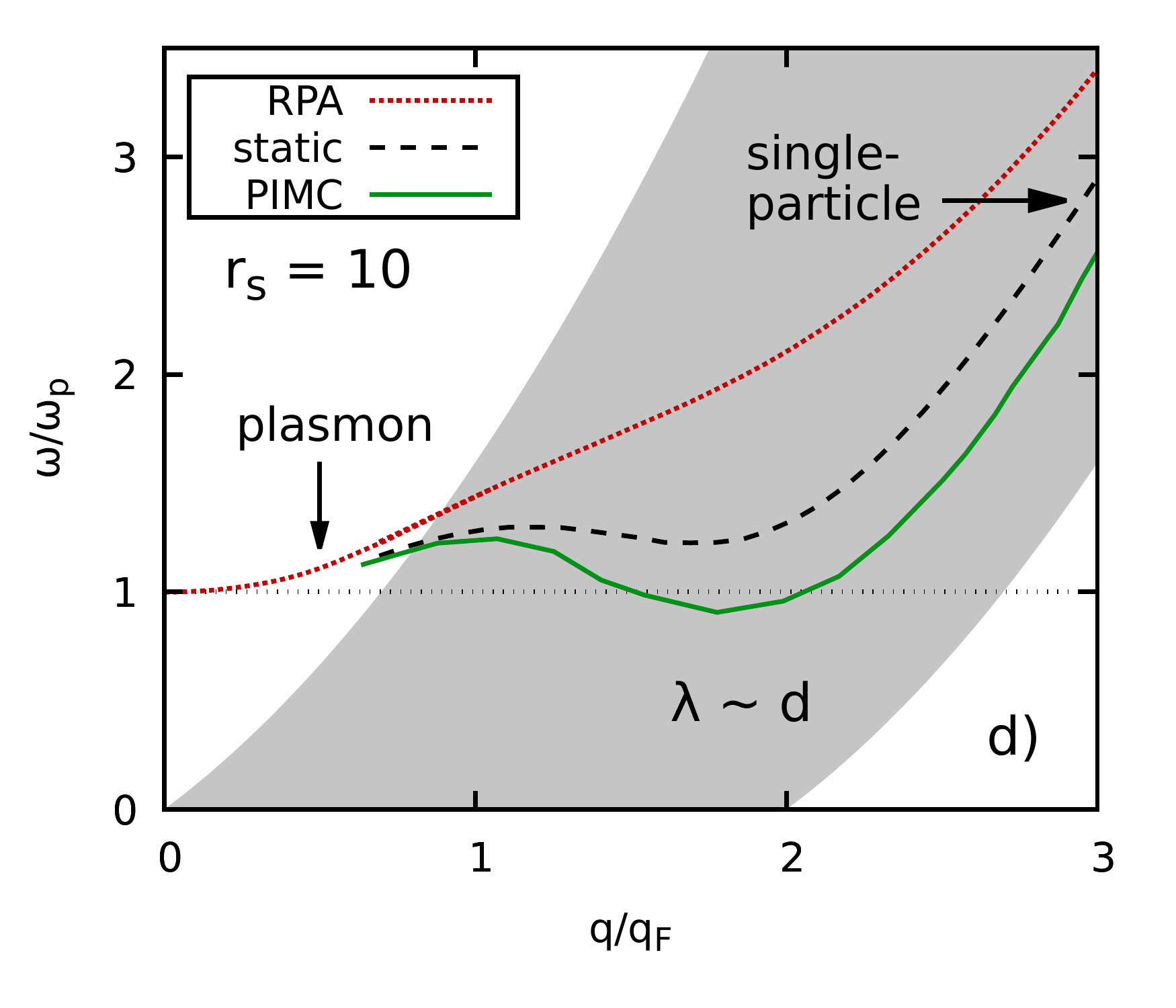}
    \caption{Top: Dynamic structure factor of the warm dense electron gas and wavenumber dispersion of its peak position for the case of weak (left) and moderate (right) coupling at the electronic Fermi temperature, $\Theta=1$. The roton feature is visible in Figure d). for the PIMC data and for the static approximation for the LFC (static). The RPA does not exhibit a non-monotonic dispersion.}
    \label{fig:dispersion_prl18}
\end{figure*}
We briefly summarize the emergence of the roton feature in the UEG as shown in Fig.~\ref{fig:dispersion_prl18}. Panels a) and b) show the DSF for the intermediate wave number $q=1.88q_\textnormal{F}$ computed within RPA (dotted red), in the \emph{static approximation} (dashed black), and using the full dynamic local field correction $G(q,\omega)$ (solid green). For $r_s=2$, the green and black curves are in perfect agreement with each other, and exhibit an exchange--correlation induced red-shift compared to the mean-field description. For $r_s=10$, the situation is considerably more interesting. Firstly, using either $G(q)$ or the full $G(q,\omega)$ leads to a substantially more pronounced red-shift with respect to the RPA curve due to the higher impact of correlations. Secondly, we observe significant deviations between the full results and the \emph{static approximation}.

The corresponding wave-number dependence of the position of the maximum in the DSF is shown in panels c) and d). Evidently, the exchange--correlation induced red-shift for $r_s=2$ is most pronounced around intermediate $q$, but no roton-type minimum occurs.
In contrast, both the \emph{static approximation} and the full PIMC solutions exhibit a roton-type feature for intermediate $q$ for $r_s=10$. In addition, we note that the minimum in $\omega_0(q)$ is more pronounced when the full $G(q,\omega)$ is used, as the \emph{static approximation} tends to merge to actual roton peak and the additional shoulder at the position of the RPA peak into a single broad feature, cf.~panel b).

From a physical perspective, the roton-type feature in the UEG at low densities has recently been explained in Ref.~\cite{dornheim_comphys_22} by the alignment of pairs of electrons. The fluctuation--dissipation theorem [Eq.~(\ref{eq:dsf-fdt})] indicates that $S(q,\omega)$ is fully described by the response of the system to an external harmonic perturbation.
If the wave length of the latter is comparable to the average interparticle distance $d$, the perturbation will induce a spatial pattern of the electrons that reduces the average interaction energy in the system. This reduction in the energy of a density fluctuation of wave number $q\sim2\pi/d$ is the root cause of the minimum in $\omega_0(q)$. Additional aspects of this pair alignment have been investigated in the recent Refs.~\cite{Dornheim_Force_2022,Dornheim_PRR_2022,Dornheim_spatial_alignment,Dornheim_insight_2022,Dornheim_PTR_2022}.


%

\subsection{Taking e-i collisions into account via the Mermin dielectric function}\label{ss:merin}
We now go beyond the assumption of a rigid ionic background (jellium) and take into account scattering of electrons with individual ions. This leads to qualitative deviations of the dielectric function from the mean field limit. The simplest approach is given by the relaxation time approximation that was introduced by Mermin \cite{mermin_prb_70} who took into account electron-ion correlation effects via a constant collision frequency $\nu$. This concept was extended to a frequency-dependent collision frequency, $\nu(\omega)$ \cite{selchow2001,reinholz2000} and combined with the description of electronic correlations using local field corrections \cite{fortmann2010}. The ``extended Mermin response function'' is expressed in terms of the jellium density response, Eq.~(\ref{eq:chi-ueg}), according to:
\begin{equation}
    \chi^{xM}(q,\omega)
    = 
    \frac{\left( 1 - \frac{i\omega}{\nu(\omega)} \right) \chi^{\rm J}\left[q,\omega+i\nu(\omega)\right]\chi^{\rm J}(q,0)}{\chi^{\rm J}[q,\omega+i\nu(\omega)]-[i\omega/\nu(\omega)]\chi^{\rm J}(q,0)}\,.
    \label{eq:chi-xm}
\end{equation}
Such an ansatz is commonly used to extrapolate density functional theory (DFT) results for the dielectric function, which is obtained in the long-wavelength limit via the Kubo-Greenwood formula. Thus  predictions for the dynamic structure factor at finite wave-vectors based on static DFT simulations become possible~\cite{plagemann2012,ramakrishna2019}.

Alternatively, the electron-ion collision frequency can be determined perturbatively from kinetic theory \cite{bonitz_98teubner,kwong_prl_00}. For the case of the uniform electron gas in equilibrium, different approximations are discussed in Ref.~\cite{Moldabekov2019}. In our calculations, we will use the following expression (labelled `RPA' in~\cite{Moldabekov2019}), which is obtained from the full Lenard-Balescu result by neglecting the plasmon feature of the dielectric function in the collision integral:
\begin{equation}\label{eq:born}
    i\nu(\omega) = \frac{\omega_p^*}{6\pi^2 n_e^* \omega} \int\limits_0^\infty \mathrm{d}q\, q^6 V_s^2(q) S_{ii}(q) \left[\epsilon(q,\omega) - \epsilon(q,0) \right]\,.
\end{equation}
Here $V_s(q)=v_q/\epsilon(q,0)$ 
is the statically screened potential. All dielectric functions are calculated in RPA. Note that unbound electrons, with a density henceforth denoted by $n_e^*$, are expected to provide the dominant contribution to screening and scattering. $S_{ii}(q)$ is the ion-ion static structure factor for which we will use  $S_{ii}(q) \approx 1$. Possible deviations from this approximation in a partially ionized hydrogen plasma will be discussed in Sec.~\ref{ss:partial-ionization}.
\begin{figure}
    \includegraphics[width=\linewidth]{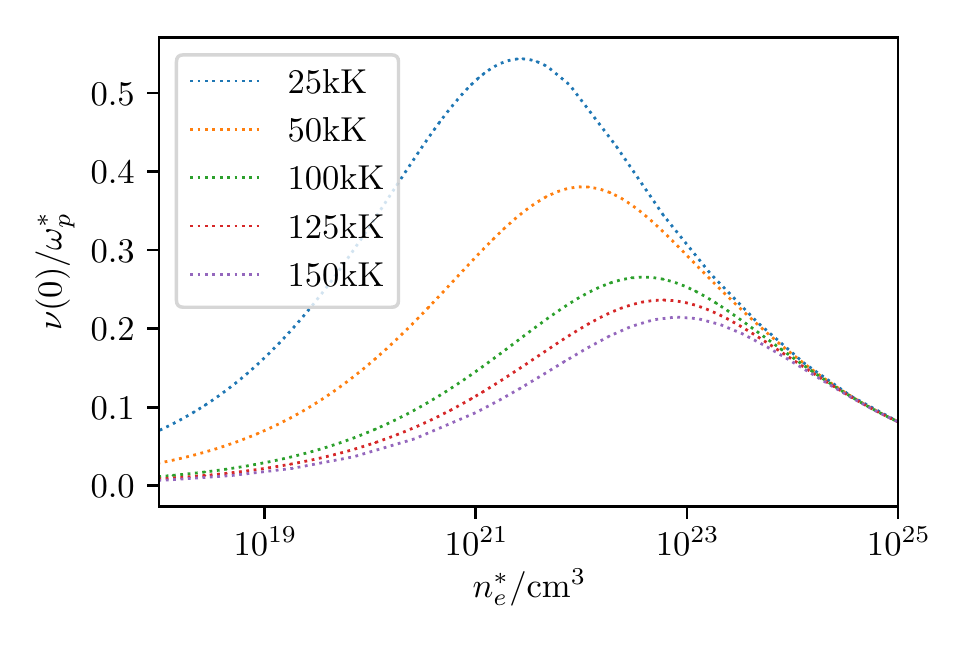}
    \caption{Dependence of the static collision frequency on the free electron density at different temperatures.}
    \label{fig:nu0}
\end{figure}
In Fig.~\ref{fig:nu0}, we present the density dependence of the static limit, $\omega \to 0$, of the collision frequency at different temperatures. 
This quantity well reflects the main trends of the effect of electron-ion collisions (the frequency dependence turns out to be of minor importance on the plasmon dispersion).
Fig.~\ref{fig:nu0} shows that collision effects decrease with the temperature as a result of decreased coupling effects. At the same time, the density dependence is non-monotonic with a maximum in the range between $10^{21}\dots 10^{23}$cm$^{-3}$. The decrease towards lower densities is due to a decrease of classical Coulomb correlation effects. On the other hand, at large densities collision effects are limited due to Pauli blocking.

\begin{figure*}
    \centering
    \includegraphics[width=\linewidth]{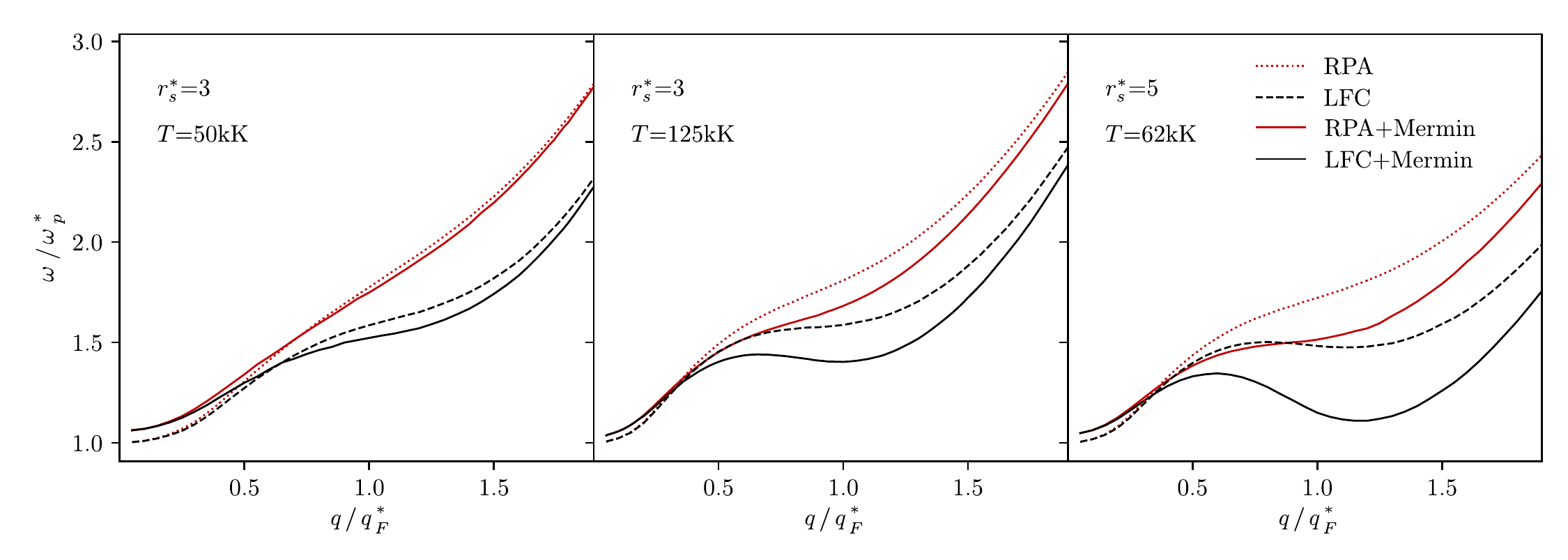}
    \caption{Dispersion of the peak of the dynamic structure factor of a two-component electron-proton plasma using the Mermin dielectric function, for different densities and temperatures. Shown are results for jellium without (RPA) and with (LFC) correlations, the two other curves are for a two-component hydrogen plasma without (Born-Mermin) and with (Born-Mermin+LFC) electron-ion collisions. The density parameter $r_s^*$ refers to the unbound electrons.}
    \label{fig:dispersion_2cp_paul}
\end{figure*}

The influence of the Mermin dielectric function on the position of the peak of the DSF and its width are presented in Figs.~\ref{fig:dispersion_2cp_paul} \& \ref{fig:dynsf-convolved}. 
It is interesting to note that both electron-electron correlations [contained in $G(q,\omega)$] and electron-ion collisions [contained in the Mermin DF] lead to an additional broadening and red shift of the peak of the DSF, compared to the RPA case which only includes Landau damping. Both correlation effects enhance one another. A striking observation is that correlations not only broaden and shift the peak of the DSF, but they also stabilize and enhance the negative dispersion  of its peak position. In a two-component plasma, this effect is predicted to appear at even lower $r^*_s$-values, i.e. higher free electron densities than for the jellium model. Since  the free electron density is reduced in a real two-component plasma due to bound state formation (atoms, molecules), the results of Fig.~\ref{fig:dispersion_2cp_paul} cannot directly be applied to a hydrogen plasma. This problem is solved in Sec.~\ref{s:hydrogen}.

\subsection{Effect of ion structure factor on e-i collision frequency\label{ss:partial-ionization}}
\begin{figure}[th]
    \centering
    \includegraphics[width=\linewidth]{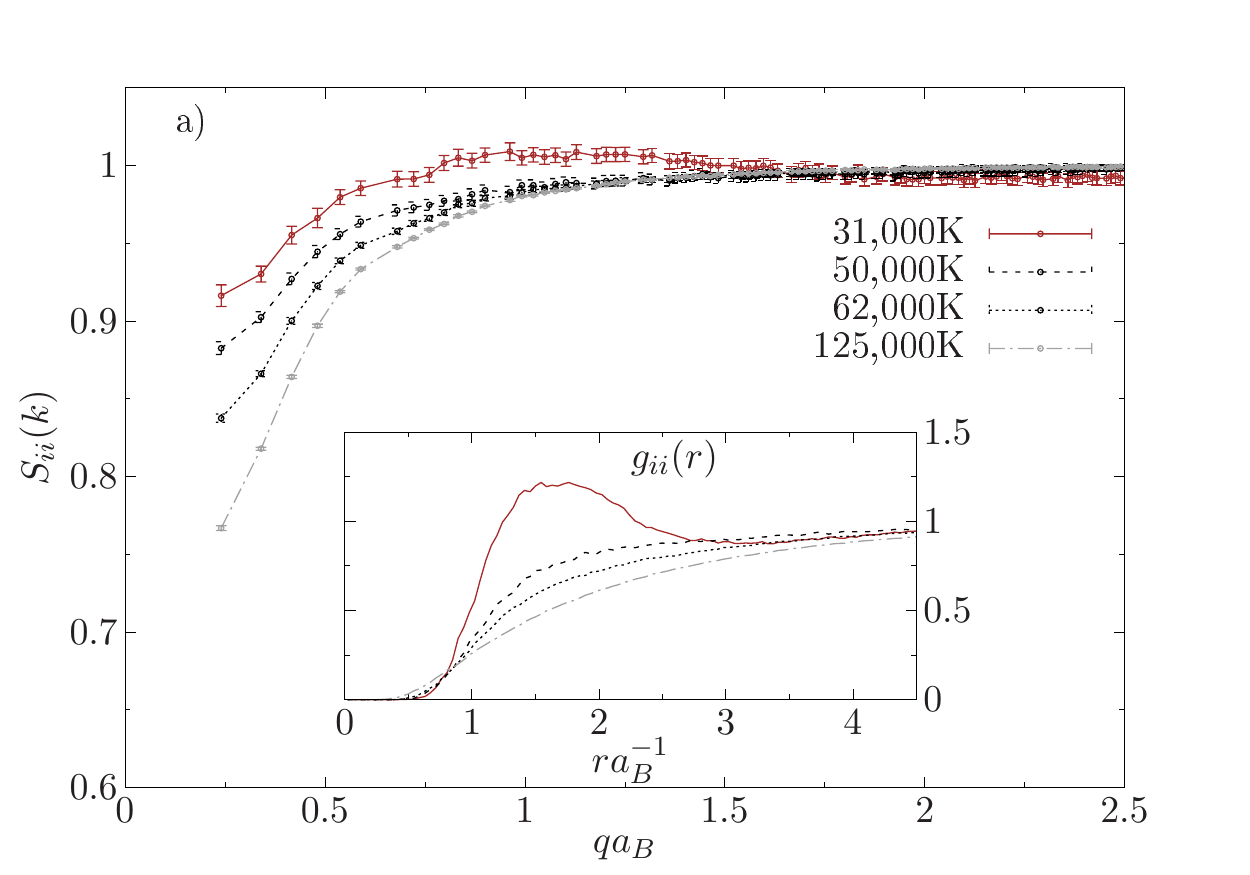}
    \caption{Static ion-ion structure factor and pair distribution function (inset) for $r_s = 5$ and four different temperatures obtained from first principles path-integral Monte Carlo simulations of a hydrogen plasma. The peak of $g_{ii}$ at the lowest temperature is due to hydrogen molecules. The Brueckner parameter $r_s^*$ that refers to the free electron density is different for each curve and can be obtained using the degree of ionization, cf. Fig.~\ref{fig:hydrogen_data}.}
    \label{fig:Sii}
\end{figure}

In the calculation of the collision frequency \eqref{eq:chi-xm}, entering the extended Mermin response function, we haveneglected the influence of the ion structure factor so far, setting $S_{ii}(q) = 1$.
To test the accuracy and validity range of this approximation we have performed direct fermionic path integral Monte Carlo simulations 
for several typical parameter combinations. We have extended the fermionic propagator approach developed for the electron gas in Ref.~\cite{filinov_cpp_21} to partially ionized hydrogen \cite{filinov_23} where, for the pair density matrix, we used the solution of the two-particle Bloch equation and the improved Kelbg potential \cite{filinov_pre04,bonitz_cpp_23}. 
 In Fig.~\ref{fig:Sii}, we show the results for four temperatures and the density parameter $r_s=5$. 
For the lowest temperature where a significant molecular fraction is observed, $S_{ii}(q)$ exhibits the largest deviations from unity. 
Recomputing the collision frequency according to Eq.~\eqref{eq:born}, we find a modification of less than $5\%$, which does not lead to noticeable modifications of the dispersion $\omega_0(q)$. The situation would change at low temperatures, where the plasma is dominated by molecules which, however, is not the region of interest for the study of electronic plasma oscillations in general, and  the roton feature in particular. This justifies the choice of $S_{ii}(q)=1$ made above.

\section{Predictions for a dense hydrogen plasma}\label{s:hydrogen} 
%
%
In this section, we explore the parameter range in which the non-monotonic $q$-dependence of the peak of the DSF occurs in a partially ionized warm dense hydrogen plasma. 
To this end, we need to define the relevant parameters in a many-component (partially ionized) plasma taking into account the reduction of the free electron number in case of bound state formation:
\begin{itemize}
   \item the degree of ionization, $\alpha=n_e^*/n_e^{\rm tot}$, with $\alpha \in [0, 1]$, with the free electron density, $n_e^*$, and the total density, $n_e^{\rm tot}=n_e^*+n_e^{\rm bound}$,
    \item the free electron Brueckner parameter \\
$ r_s^* = \overline{a}^*/{a_B}$,    where $\overline{a}^* = [3/(4 \pi n_e^*)]^{1/3}$ is the Wigner-Seitz radius corresponding to the free electrons; the relation to the standard Brueckner parameter is $r_s^*=\alpha^{-1/3}\cdot r_s \ge r_s$,
    \item the dimensionless wavenumber,\\ $\bar q^* = q/q^*_F=\alpha^{-1/3}\cdot \bar q$, where $q^*_F = (3 \pi^2 n_e^*)^{1/3} = \alpha^{1/3}\cdot q_F \le q_F$ is the Fermi wave number of the free electrons
    \item the dimensionless temperature of the free electrons, $\Theta^* = k_B T/ E^*_F=\alpha^{-2/3}\cdot \Theta \ge \Theta$, with the Fermi energy of the free electrons, $E^*_F=\frac{\hbar^2 q_F^{*2}}{2m}=\alpha^{2/3}\cdot E_F \le E_F$,
    \item the plasma frequency of the free electrons,\\
$ \omega^*_p =  [(n^*_e e^2)/(\epsilon_0 m_e)  ]^{1/2}$. 
%
In atomic units we have $ \omega^*_p =\sqrt{3/(r_s^*)^3} = \alpha^{1/2}\cdot \omega_p \le \omega_p$.
\end{itemize}

%
We now translate the range of dimensionless parameters  
of the negative plasmon dispersion of the UEG model to the corresponding density and temperature range for warm dense hydrogen. As discussed above, the plasmon dispersion is due to the oscillation of the free electrons with density $n^*_e$. This means, a first rough estimate for the density and temperature range where the non-monotonic dispersion is predicted in hydrogen can be obtained from Fig.~\ref{fig:phase_diagram} by replacing, on the axes, $r_s \to r_s^*$ and $\Theta \to \Theta^*$.

\begin{figure}
    \centering
    \includegraphics[width=0.5\textwidth]{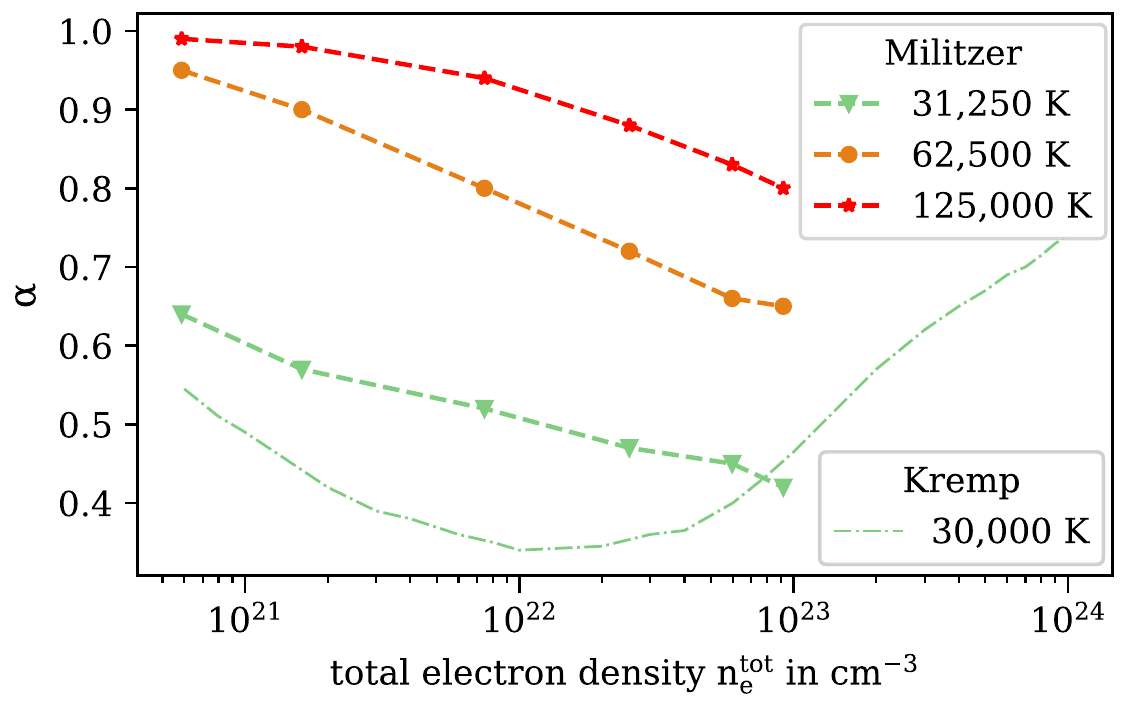}
    \caption{Selected isotherms of the degree of ionization of hydrogen from B. Militzer et al. \cite{2001Militzer}. For comparison the 30,000 K isotherm from D. Kremp \textit{et al.}  \cite{QStatNonideal} was added. At the largest density, $n_e^{\text{tot}} \approx 10^{24} \text{cm}^{-3}$, corresponding to $r_s \approx 1.2$, the plasma is expected to be nearly fully ionized \cite{bonitz_prl_5}. In addition, full ionization is also expected at temperatures above  150,000 K. }
    \label{fig:hydrogen_data}
\end{figure}

First, from $r_s^*$ and $\Theta^*$ the free electron density, $n^*_e$, and the temperature, $T$, are calculated. However, the temperature dependence of $\Theta^* $ is in general only valid for free electrons. Therefore, this has to be converted to total electron density. To this end, we use isotherms of the degree of ionization $\alpha(n^{\rm tot},T)$, see Fig.~\ref{fig:hydrogen_data}. 
\begin{figure}
    \centering
    \includegraphics[width=0.5\textwidth]{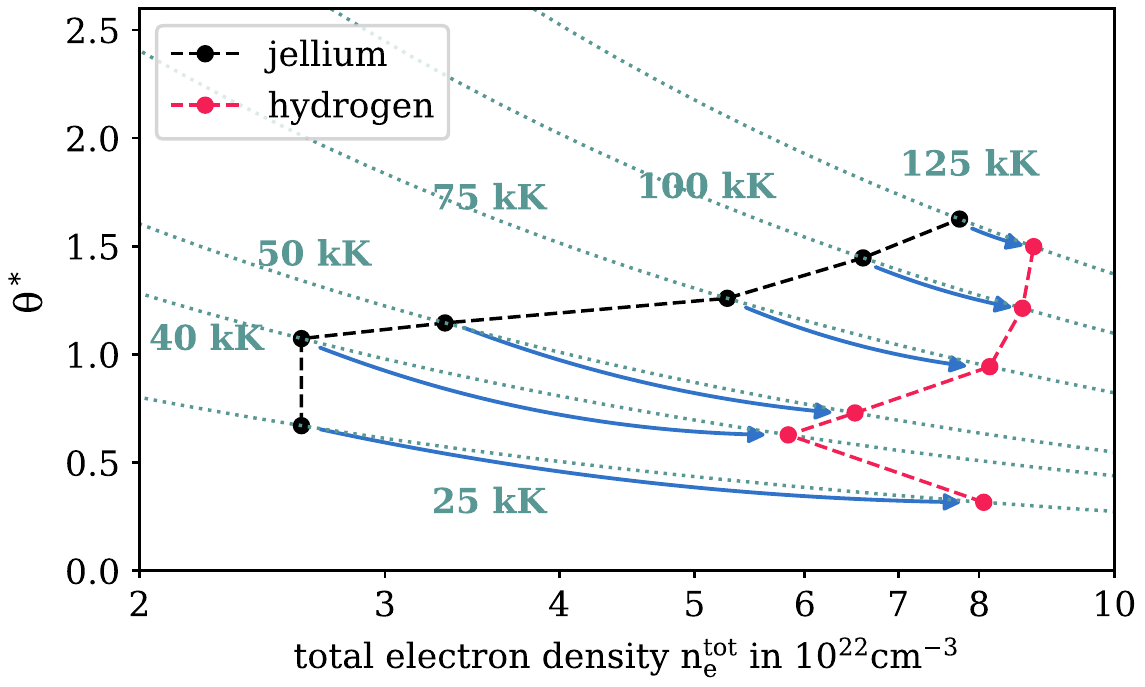}
    \caption{Illustration of the density shift along isotherms, from a uniform electron gas (black curve: fully ionized plasma with density $n_e^*$, Mermin results) to a partially ionized hydrogen plasma (red) with total electron density $n_e^{\text{tot}} = n_e^* + n_e^{\text{bound}}$ The vertical axis is scaled to the Fermi temperature of the free electron component, $\Theta^*=k_BT/E_F^*$. 
}
    \label{fig:densShift}
\end{figure}
\begin{figure}
    \centering
    \includegraphics[width=0.5\textwidth]{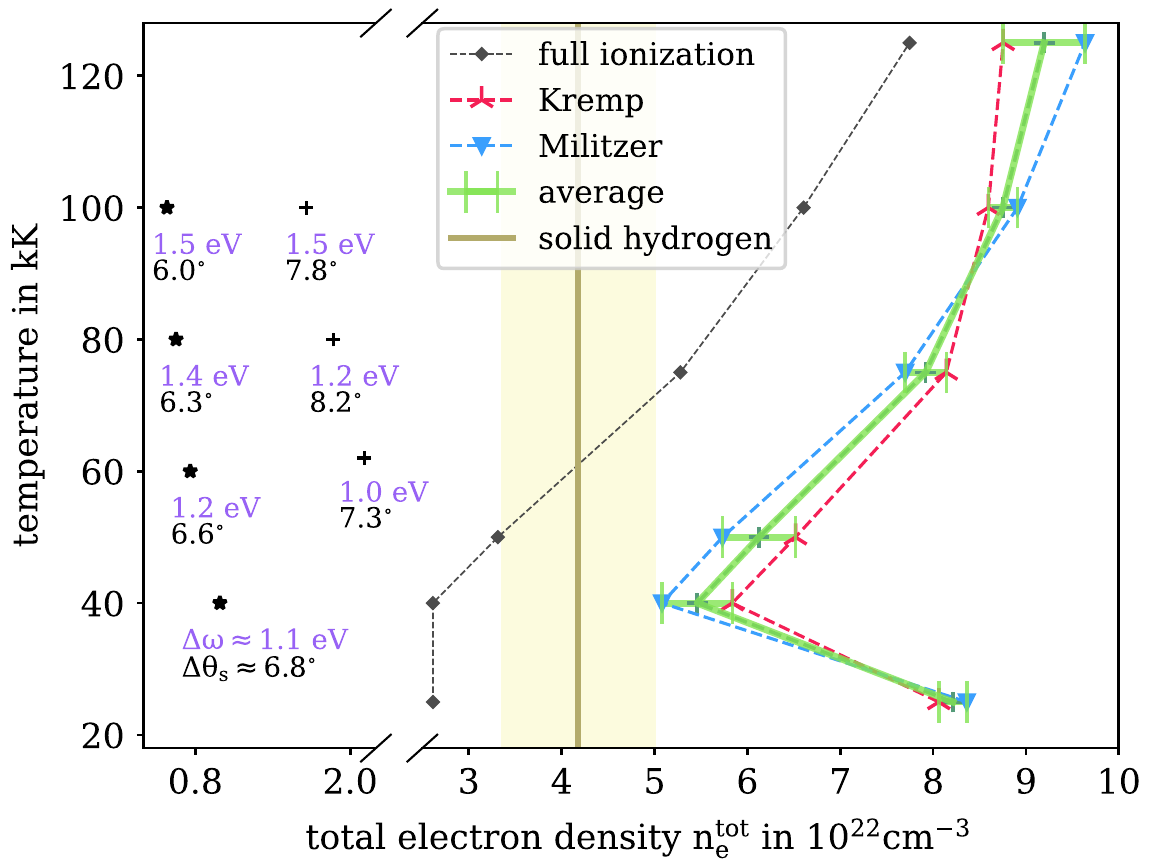}
    \caption{``Phase diagram'' of the negative plasmon dispersion for jellium and hydrogen. Negative dispersion is predicted to exist to the left of the curves. Black curve: Mermin results for jellium (full ionization). Colored curves: 
results for partially ionized hydrogen plasma with
the degree of ionization taken either from 
D. Kremp \textit{et al.}  \cite{QStatNonideal} (red), or from B. Militzer et al. \cite{2001Militzer} (blue). The green curve is the arithmetic mean of the red and blue curves. The yellow stripe corresponds to the parameters accessible with hydrogen jets around the density of solid hydrogen (vertical line) $\pm 20\%$   \cite{2022Fletcher}.
Symbols in the left part refer to data points listed in Tab.~\ref{tab:observe_roton}: ``+'' refers to $r_s^*=5$ and ``$*$'' to $r_s^*=7$. Numbers next to the symbol correspond to $\Delta\omega$ (in eV) and $\Delta\theta_s$ (for a photon energy of $6$\,keV, cf. Fig.~\ref{fig:scatt-angles}).}
    \label{fig:resultdensity}
\end{figure}
From these data sets, the free electron number density $n_e^*(n,T) = \alpha(n,T) \cdot n$ can be calculated for all data points. 
Thus, a specific free electron density at a given temperature value is mapped to the total electron density in a unique way, as demonstrated in Fig.~\ref{fig:densShift}. Because there are data points at temperatures without any available hydrogen data, a linear interpolation is used to calculate the  isotherms for each of the data points in Fig.~\ref{fig:phase_diagram}. 
The final result of the conversion is given in Fig.~\ref{fig:resultdensity}.

\subsection{Experimental wave number range}\label{ss:q-experiment}
\begin{figure}
    \centering
    \includegraphics[width=0.5\textwidth]{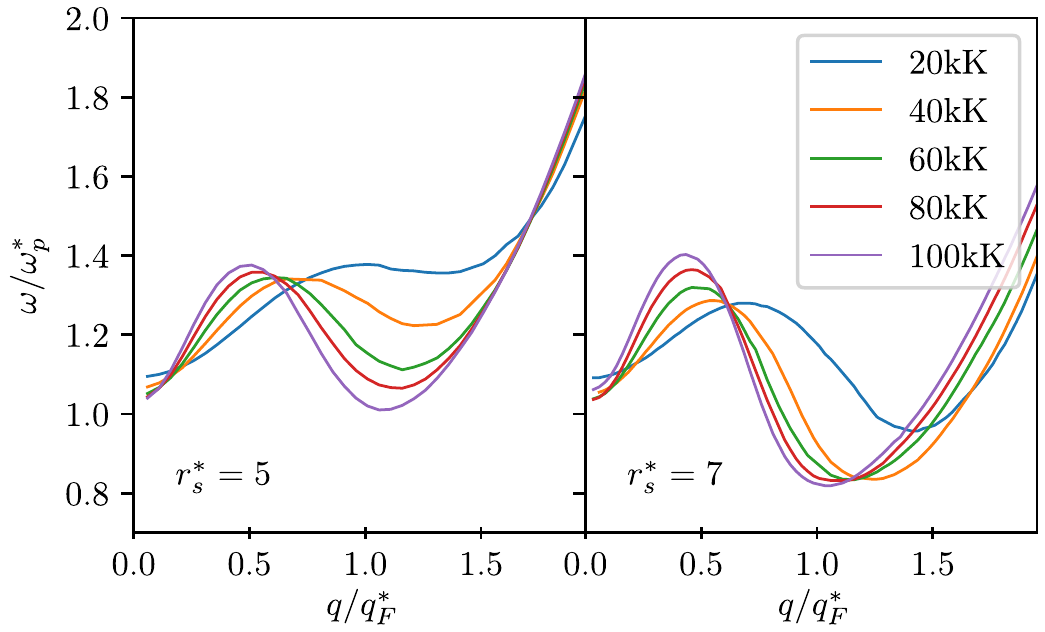}
    \caption{Mermin+LFC results for the peak position of the dynamic structure factor vs. wavenumber. Shown are results for two densities and five temperatures each.}
    \label{fig:dispersion-paul}
\end{figure}
We now analyze the photon energy and scattering angles that are suitable to detect the roton feature in Thomson scattering experiments with dense hydrogen, for the illustration and notation see Fig.~\ref{fig:dispersion_schematic}. Let us first consider the wave number range for jellium.  The example of $r_s^* = 10$ and $\Theta^* = 1.0$ is shown in Fig.~\ref{fig:dispersion_prl18}
for which the wave numbers bracketing the negative dispersion $d\omega_0/dq < 0$ are given by
~\cite{dornheim_prl_18} $q_1=1.1 q_F^*$ and $q_{\rm min}=1.88 q_F^*$.
For hydrogen the wave numbers shift considerably, as is shown in Fig.~\ref{fig:dispersion-paul}  where data for values of $r_s^*=5$ and $r_s^*=7$ are presented. These data are obtained from the Mermin dielectric function with static LFC included, as explained in Fig.~\ref{fig:dispersion_2cp_paul}.

\begin{figure}
    \centering
       \includegraphics[width=0.45\textwidth]{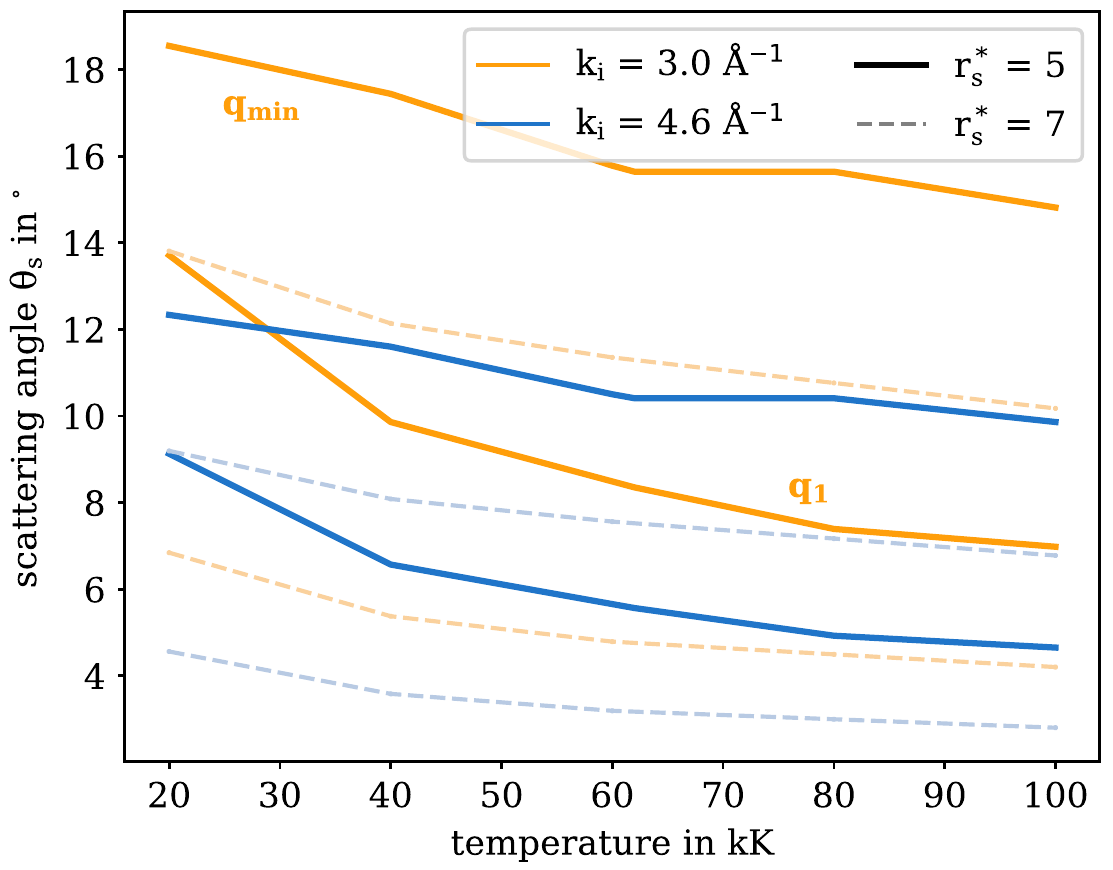}
    \caption{
    Temperature dependence of the Thomson scattering angles required to detect the roton feature in hydrogen for two photon energies $k_i = 3 \text{\AA}^{-1}$ (6 keV, orange) and  $ k_i=4.6 \text{\AA}^{-1}$ (9 keV, blue) and two densities $r_s^* = 5$ ($\approx 1.29 \cdot 10^{22} \text{cm}^{-3}$, solid lines) and $r_s^* = 7$ ($\approx 0.47 \cdot 10^{22} \text{cm}^{-3}$, dashed lines). Upper (lower) lines of each pair of lines with the same color and style correspond to the minimum (maximum) of the dispersion $\omega_0(q)$, cf. Fig.~\ref{fig:dispersion-paul}. The angles were computed using Eq.~(\ref{eq:angle}).}
    \label{fig:scatt-angles}
\end{figure}

An approximation for the absolute value of the momentum transfer $\hbar q$ dependent on the scattering angle $\theta_s$ in Thomson scattering is~\cite{Dornheim_T2_2022} 
\begin{equation}\label{eq:angle}
q \approx 2 k_i \mathrm{sin(\theta_s/2)},
\end{equation}
where $k_i = \frac{2 \pi E_i}{h c}$ is the incident laser wave number for an initial photon energy $E_i$. 
Using typical photon energies of $\hbar \omega = (6\dots 9) \,\mathrm{keV}$ that are used in XRTS experiments at free electron lasers, scattering signals for the two wave numbers should be detectable simultaneously using two detectors placed under different angles $\theta_s$. Then, using Eq.~\eqref{eq:angle} we can compute the angles for different temperatures. The results are shown in Fig.~\ref{fig:scatt-angles}.

\begin{table}[h]
  \centering
\begin{tabular}{cccccccc} 
\multicolumn{8}{c}{$r_s^* = 5$ ($n_e^* \approx 1.29 \cdot 10^{22} \text{cm}^{-3}$)} \\
\cline{1-8}
  \makecell{T \\ in kK }  &  \makecell{$q_1$ \\ in $q_F^*$   } &   \makecell{$q_{\rm min}$ \\ in $q_F^*$ } &   \makecell{$\omega_{\text{max}}$ \\ in $\omega_p^*$ } &  \makecell{$\omega_{\text{min}}$ \\ in $\omega_p^*$ }   &  
  \makecell{$\Delta \omega^*$ \\in $\omega_p^*$}   & \makecell{$\Delta \omega$\\in eV}  &   \makecell{$n_e^{\text{tot}}$ in\\$10^{22} \text{cm}^{-3}$ }    \\ \hline 
    \\
20&1.00&1.35&1.38&1.36&0.02&0.08&6.24\\  
40&0.72&1.27&1.34&1.22&0.12&0.51&2.75\\  
60&0.62&1.15&1.34&1.11&0.23&0.97&2.15\\  
62&0.61&1.14&1.35&1.11&0.24&1.01&2.11\\  
80&0.54&1.14&1.36&1.07&0.29&1.22&1.87\\  
100&0.51&1.08&1.37&1.01&0.36&1.52&1.66\\ 
\\
\multicolumn{8}{c}{$r_s^* = 7$ ($n_e^* \approx 0.47 \cdot 10^{22} \text{cm}^{-3}$)} \\
\cline{1-8}
  \makecell{T \\ in kK }  &  \makecell{$q_1$ \\ in $q_F^*$   } &   \makecell{$q_{\rm min}$ \\ in $q_F^*$ } &   \makecell{$\omega_{\text{max}}$ \\ in $\omega_p^*$ } &  \makecell{$\omega_{\text{min}}$ \\ in $\omega_p^*$ }   &  
  \makecell{$\Delta \omega^*$ \\in $\omega_p^*$}   & \makecell{$\Delta \omega$\\in eV}  &   \makecell{$n_e^{\text{tot}}$ in\\$10^{22} \text{cm}^{-3}$ }    \\ \hline 
    \\
20&0.70&1.41&1.28&0.96&0.32&0.81&2.93\\  
40&0.55&1.24&1.29&0.84&0.45&1.14&0.99\\  
60&0.49&1.16&1.32&0.84&0.48&1.22&0.76\\  
80&0.46&1.10&1.36&0.83&0.53&1.35&0.65\\  
100&0.43&1.04&1.40&0.82&0.58&1.47&0.58\\ 
\end{tabular}
  \caption{Parameters of the roton feature: minimum (maximum) of the wave number, $q_{\rm min}$ ($q_1$) and frequency, $\omega_{\text{min}}$ ($\omega_{\text{max}}$), and the depth of the  minimum, $\Delta \omega^*$ [in units of $\omega_p^*$] and $\Delta \omega$ [in eV] for two electron densities $r_s^* = 5$ and $7$, for five temperatures. The data refer to Fig.~\ref{fig:dispersion-paul}. In addition, the total electron density, $n_e^{\text{tot}}$, is calculated with the procedure from Sec.~\ref{s:hydrogen}. }
  \label{tab:observe_roton}
\end{table}

\subsection{Prospects for observing the roton feature in hydrogen experiments}\label{ss:instrument-function}
After analyzing in Fig.~\ref{fig:dispersion-paul} and table~\ref{tab:observe_roton} the expected relevant wave number range $[q_1,q_{\rm min}]$  and frequency change $\Delta \omega$ for the roton feature, we now discuss the prospects for an experimental observation.

\begin{figure}
    \centering
    \includegraphics[width=0.5\textwidth]{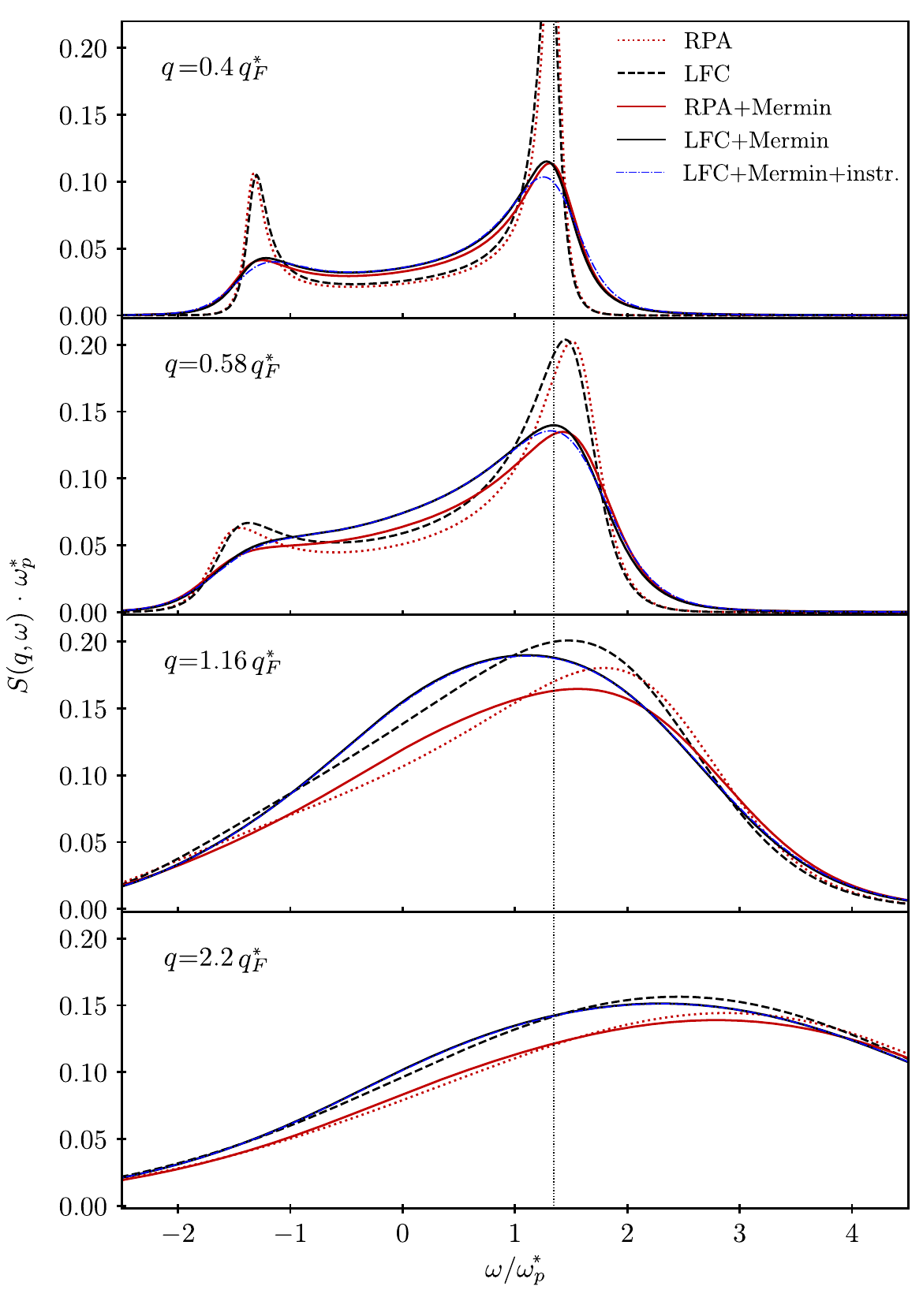}
    \caption{Dynamic structure factor at $r_s^*=5, T=62,000K$, for different approximations, as explained in the inset. Blue curve: Mermin+LFC result which is additionally convolved with a Gaussian instrument function with $\sigma = 3.65 eV \approx 0.1378 \omega_p^*$.
    The vertical dotted line indicates the local maximum $\omega_{\rm max}$ of the peak position (Mermin+LFC). }
    \label{fig:dynsf-convolved}
\end{figure}

In order to do so, a finite resolution of the frequency measurement needs to be taken into account. This is analyzed in Fig.~\ref{fig:dynsf-convolved} where a Gaussian instrument function with a realistic width of $\sigma=3.65$eV is used. %
The figure shows that, for typical densities and temperatures, this broadening has only a very small effect on the dynamic structure factor and does not change the dispersion $\omega_0(q)$ significantly. This conclusion does not change if the width is further increased by a factor $2$ to $3$.

We analyze the data presented in Tab.~\ref{tab:observe_roton} to find the optimal parameters to detect the roton feature experimentally. For jellium, the minimum in the dispersion, i.e. $\Delta\omega/\omega_{pl}$ is more pronounced for increased temperatures and  for larger values of 
 $r_s^*$ (smaller free electron densities), see Fig.~\ref{fig:dispersion-paul}. 
 For the parameter range studied in Fig.~\ref{fig:dispersion-paul} $\Delta\omega/\omega_{pl}$ is maximal at 100,000 K, for both values of $r_s^*$, cf. Tab.~\ref{tab:observe_roton} and is larger for $r_s^*=7$.
 
 For the case of partially ionized hydrogen, we now convert the frequency difference into absolute units (eV), $\Delta\omega^* \to \Delta\omega$, as done in Tab.~\ref{tab:observe_roton}.
As before, at a given free-electron density, $\Delta\omega$ increases with temperature. However, for a given temperature, the density dependence becomes more complex. In particular, for the highest temperatures, the frequency difference $\Delta\omega$ only weakly depends on $r_s^*$.
For illustration, we included all points from the table for which $\Delta\omega$ exceeds $1$eV in Fig.~\ref{fig:resultdensity}.

Equally important for the experimental setup is the difference of the  scattering angles, $\Delta\theta_s$, that refer to $q_1$ and $q_{\rm min}$, respectively, cf. Fig.~\ref{fig:scatt-angles}. For all temperatures, $\Delta\theta_s$
is largest at the smaller photon energy of 6 keV. 
Also, $\Delta\theta_s$ increases with the free electron density, even though this effect is less pronounced. The largest value, $\Delta\theta_s\approx 8.25^{\circ}$, for a mean value of $\theta_s\approx 11.5^{\circ}$, is obtained for $r_s^* = 5$,  at 80,000 K. 
For the lower density, $r_s^* = 7$, the difference between angles is smaller. A further decrease of $\Delta\theta_s$ is observed for
the higher photon energy of 9 keV, for $r_s^* = 5$ and $r_s^* = 7$, in this order. The only exception is at 20,000 K, where $\Delta\theta_s$ is larger for $r_s^* = 7$ than for $r_s^* = 5$. The angle difference $\Delta\theta_s$ for these two values of $r^*_s$ and several temperature are included in Fig.~\ref{fig:resultdensity} together with the frequency change $\Delta\omega$.

We can give give a simple estimate for the angle difference $\Delta\theta_s$ using Eq.~\eqref{eq:angle} where, for the present small angles, the sine can be replaced by its argument. Furthermore, using
    $q_F \approx 3.63 \cdot (r_s^*)^{-1} \text{\AA}^{-1}$,
we obtain for the angle difference in degrees
\begin{align}
    \Delta\theta_s[^\circ] \approx \frac{\bar q_{\rm min}-\bar q_1}{k_i[\AA^{-1}]} \cdot \frac{207.9}{r_s\, \alpha[n_e^{\rm tot},T]}\,,
    \label{eq:delta-theta}
\end{align}
where $\bar q=q/q^*_F$, and we used $r_s^*=r_s \alpha$. Thus the angle difference increases if the total density is increased ($r_s$ is lowered) and if the degree of ionization decreases.
Reading off the values of $\bar q_{\rm min}$ and $\bar q_1$, from Fig.~\ref{fig:dispersion-paul}, formula (\ref{eq:delta-theta}) yields results for $\Delta\theta_s$ that are in very good agreement with Fig.~\ref{fig:scatt-angles}. 

%



 A successful experimental determination of the dispersion  depends on the obtainable precision when determining the peak positions. In recent XRTS experiments, the plasmon shift has been used to determine the free electron density, e.g., $n_e^* = 2.5 \cdot 10^{22} \text{cm}^{-3} \pm 16 \%$ from the plasmon shift of $7$ eV (deuterium \cite{2016Davis}), or $n_e^* = 1.8 \cdot 10^{23} \text{cm}^{-3} \pm 5 \%$ from the plasmon shift of $19$ eV (aluminum \cite{2013Fletcher, 2016GlenzerLCLSWDM}). This corresponds to uncertainties in energy in the range of $1-2$ eV in the best fit of the applied theoretical model.
 
%
The resolution in XRTS is currently mainly limited by the X-ray bandwidth and the resolution of the spectrometer. The limitations of the SASE bandwidth can be overcome by using seeded X-rays, especially self-seeded X-rays \cite{2016GlenzerLCLSWDM,2017ReviewFEL,2019SelfSeeding,2019SynchBook}. 
 For example, Ref.~\cite{2017ReviewFEL} reports the SASE bandwidth to be on the order of $0.1 \dots 0.2 \%$ and the seeded bandwidth to be around $0.005 \dots 0.01 \%$. Thus, the band width could be on the order of 0.4 eV for a photon energy of 8 keV which is in the range of $\Delta \omega$, see above. However, the precision is essentially limited by the resolution of the spectrometer, which is at best around $0.1 \%$ \cite{2016GlenzerLCLSWDM}, i.e., for the 8 keV photon energy the spectrometer resolution would be 8 eV. 

%
Up to now, the highest measured energy resolution of a few meV with X-ray bandwidth $10^{-4} \%$ was demonstrated in inelastic X-ray scattering measurements of the phonon dispersion of single-crystal diamond at room temperature and $T\sim 500$K at European XFEL in Ref.~\cite{Descamps_sci-rep_2020}. With the originally intended application of seeded X-rays, there will be even further improvements \cite{2018McBride}.

\section{Discussion}\label{s:discussion}
In this paper, we analyzed whether an exciting correlation effect -- the roton feature -- that was observed in a variety of systems including superfluid helium and cold alkali metals \cite{quantum_theory}, 
may also show up in warm dense hydrogen. The motivation was that this feature was recently predicted to exist also in the model of the warm dense electron gas \cite{dornheim_prl_18} and was explained to be due to the spatial alignment of pairs of electrons for certain densities \cite{dornheim_comphys_22}. 
Being carried by the free electrons, plasma oscillations are routinely observed in dense partially ionized plasmas where they are detected by X-ray Thomson scattering and serve as an important diagnostic for the plasma parameters. Therefore, it is tempting to inquire whether the roton feature will feature in dense plasmas too, and under what conditions.
The results of the present paper provide strong confirmation for this effect to be observable in dense hydrogen.

Our analysis was based on path integral Monte Carlo simulations for the strongly coupled electron component that were combined with the Mermin formalism to compute the electron response function of the two-component electron-proton plasma. The presence of the ion component does not destroy the roton feature. In contrast, this feature is appears even stronger compared to the uniform electron gas. It is stabilized and extends towards higher densities. By taking into account the partial ionization of hydrogen, we estimated the total electron densities and temperatures for which the effect should be observable in XRTS experiments with hydrogen. Good candidates are states of solid or liquid hydrogen produced by jets that undergo moderate expansion. We specified the experimental resolution necessary to observe the effect. First, the frequency resolution necessary to detect the monotonic behavior of $\omega_0(q)$ has to be in the range of 1eV. Second, the difference in scattering angles that has to be resolved to probe the relevant points on the dispersion curve is around 8 degrees. These parameters pose a challenge to current experiments but should be well within range of upcoming XRTS measurements at X-ray free electron lasers.

\section*{Acknowledgments}
We acknowledge helpful comments from Hanno Kählert. This work is supported by the Deutsche Forschungsgemeinschaft via project BO1366-13/2.
This work was partly funded by the Center for Advanced Systems Understanding (CASUS) which is financed by Germany's Federal Ministry of Education and Research (BMBF) and by the Saxon Ministry for Science, Culture and Tourism (SMWK) with tax funds on the basis of the budget approved by the Saxon State Parliament.
The PIMC calculations were carried out at the Norddeutscher Verbund f\"ur Hoch- und H\"ochstleistungsrechnen (HLRN) under grant shp00026, and on a Bull Cluster at the Center for Information Services and High Performance Computing (ZIH) at Technische Universit\"at Dresden.

\bibliography{bibliography,bibliographyLinda,mb-ref,bib-paul}
\end{document}